\documentclass[aps,prd,onecolumn,superscriptaddress,nofootinbib]{revtex4}
\usepackage{amsmath, amsfonts,bbm}
\usepackage{comment}
\usepackage[colorlinks=true,hyperfootnotes=true,citecolor=cyan]{hyperref}
\usepackage[utf8]{inputenc}

\usepackage{xcolor}

\newcommand{\TT}{T}
\newcommand{\mS}{{\mathcal S}}
\newcommand{\dd}{{\rm d}}
\newcommand{\ad}{{\rm ad}}

\newcommand{\be}{\begin{equation}} 
\newcommand{\ee}{\end{equation}}
\newcommand{\bea}{\begin{equation}\begin{aligned}}
\newcommand{\eea}{\end{aligned}\end{equation}}

\newcommand{\mpl}{M_{\rm Pl}}

\newcommand{\R}[1]{}
\newcommand{\B}[1]{}

\begin{document}

\title{\bf Minkowski space in $f(T)$ gravity}

\author{Jose Beltr{\'a}n Jim\'enez}
\email{jose.beltran@usal.es}
\affiliation{Departamento de F\'isica Fundamental and IUFFyM,Universidad de Salamanca, E-37008 Salamanca, Spain}
\author{Alexey Golovnev}
\email{agolovnev@yandex.ru}
\affiliation{Centre for Theoretical Physics, The British University in Egypt, 11837 El Sherouk City, Cairo Governorate, Egypt}
\author{Tomi Koivisto}
\email{tomik@astro.uio.no}
\affiliation{Laboratory of Theoretical Physics, Institute of Physics, University of Tartu, W. Ostwaldi 1, 50411 Tartu, Estonia}
\affiliation{National Institute of Chemical Physics and Biophysics, Rävala pst. 10, 10143 Tallinn, Estonia}
\author{ Hardi Veerm{\"a}e}
\email{hardi.veermae@cern.ch}
\affiliation{National Institute of Chemical Physics and Biophysics, Rävala pst. 10, 10143 Tallinn, Estonia}

\begin{abstract}

The full set of solutions of $f(T)$ gravity with the Minkowski metric is considered in this note. 
At the 4-th order in perturbations around the trivial tetrad solution, a new mode is found explicitly.  Its presence signals a strong coupling problem that transcends the Minkowski background and also suggests the pathological nature of cosmological solutions.

\end{abstract}

\maketitle

\section{Introduction}

General Relativity describes the interaction of matter fields mediated by a self-interacting massless spin-2 field, the graviton. Distinctive to the gravitational interaction is its geometrical interpretation admitted by the equivalence principle. While gravity is thus related to the curvature of the spacetime metric, there are equivalent formulations of General Relativity in terms of flat connections \cite{BeltranJimenez:2019tjy, Jimenez:2019ghw}. The most extensively studied of these reformulations imposes the metric-compatibility of the flat connection and is known as the (metric) Teleparallel Equivalent of General Relativity \cite{Aldrovandi:2013wha} (TEGR). In a vast number of studies the Lagrangian of TEGR, defined by the so called torsion scalar $T$, has been promoted to an
arbitrary nonlinear function $f(T)$. In this note we report a new result which may shed light on the currently undisclosed degrees of freedom (dof's) in the modified $f(T)$ gravity.  

An important ingredient of the TEGR Lagrangian is that the torsion scalar $T$ features a local Lorentz symmetry realised up to a total derivative, irrelevant for the classical field equations (relevant for quantum corrections though). It is precisely this total derivative generated by the torsion scalar under local Lorentz transformations that  crucially breaks the local Lorentz symmetry for the non-linear extensions conforming the $f(T)$ theories \cite{Ferraro:2006jd,Li:2010cg}, though they retain a global Lorentz invariance. The most important consequence of the breaking of the local Lorentz symmetry is the appearance of new dof's. The nature and even the number of the new dof's within $f(T)$ gravity remains as one of its most important and puzzling issues~\cite{Li:2011rn, Ferraro:2014owa, Cai:2015emx,  Nester:2017wau, Ferraro:2018tpu, Ferraro:2018axk, Bejarano:2019fii, Golovnev:2019kcf, Ferraro:2020tqk}. There is very little doubt about the existence of at least one of them, in part strongly supported by the reduced local symmetries of the theories. These additional dof's however appear very elusive since they have never been explicitly seen in perturbation theory around a given background. In particular, they did not show up in the theory of linear cosmological perturbations \cite{Golovnev:2018wbh}. A full Hamiltonian analysis seems to however indicate the existence of one extra dof and, if confirmed, this would indeed signal towards a strong coupling problem of any cosmological background, including Minkowski and (anti) de Sitter backgrounds.

Unlike the case of $f(R)$, the new modes in $f(T)$ are not associated with higher derivatives, but they come due to the local Lorentz violation in the space of tetrads (in pure tetrad formalism). One can reformulate this statement by saying that the extra dof's are responsible for families of inequivalent solutions with one and the same metric. 
Therefore, a possible approach to the issue of the number of degrees of freedom would be to study all solutions of $f(T)$ gravity with a given simple metric. The simplest choice is, of course, Minkowski spacetime. In this paper we undertake such an analysis considering the simplified problem of the dynamical evolution of the tetrad with a flat metric. Our analysis implies the existence a new mode and gives the energy scale at which it is expected to show up. Our findings do not contradict the claim of Ref.~\cite{Ferraro:2018tpu} that only one extra mode exists.

\section{Flat space in $f(\TT)$ and extra modes}

Due to the local Lorentz breaking, solutions of $f(\TT)$ equations of motion are not completely specified by the metric. Instead, we need to solve for the tetrad, and generically, for an admissible metric, there are many of them, though not all solve the equations of motion. In particular, one can look for the class of vacuum solutions with Minkowski metric. Of course, the simplest choice for this metric is the trivial tetrad $e^a_{\mu}=\delta^a_{\mu}$. The generic tetrad reads
\be
\label{flattetrad}
	e^a{}_{\mu}=\Lambda^a{}_b \delta^b{}_{\mu}
\ee
where $\Lambda^a_b (x)$ is an arbitrary smooth function on the space-time which takes values in $SO(1,3)$. At the linearised (around the trivial tetrad) level, any such tetrad is a solution of vacuum $f(T)$ equations of motion. This is due to the accidental restoration of the full local Lorentz invariance at the level of quadratic action around the trivial tetrad. However, non-perturbatively it is certainly not the case, and our aim is to describe the subclass of flat tetrads \eqref{flattetrad} which solve the equations.

In an exact treatment, the complete set equations of motion, which also include perturbations of the metric, should be considered. However, we can consider a simplified procedure, which, as we will shortly show, is expected to hold in the decoupling limit. Namely, let us substitute the tetrad \eqref{flattetrad} into the action
\be
\label{action}
	\mS=\frac12\int \dd^4 x \|e\| \cdot f(\TT)
\ee
of $f(T)$ gravity. Since in this case $0=\TT+2\partial_{\mu}T^{\mu}$, it follows that
\be
\label{T}
	{\TT}
	= 2\delta^{\nu}_a \delta^b_{\mu}\partial^{\mu}\left(\Lambda^{-1} \partial_{\nu} \Lambda\right)^{a}{}_{b}.
\ee
The action for the extra modes is given by substituting relation \eqref{T} into the action \eqref{action}.  Around the flat metric solutions, stationarity of this restricted action is a necessary condition for stationarity of the full $f(T)$ action under arbitrary small variations of the tetrad. In essence, we only consider the antisymmetric part of equation of motion. In this way, the simplified system should give an upper bound for the number of extra degrees of freedom.

The procedure outlined above can be justified from the vantage point of the covariant formulation~\cite{Golovnev:2017dox}. Restoring the physical dimensions (in natural units) the action must be expressed as
\be
\label{Maction}
	\mS=\frac{1}{2}M_P^2 M^2\cdot\int \dd^4 x \|e\| \cdot f\left(\frac\TT{M^2}\right).
\ee
It contains two characteristic scales, the Planck scale $M_P$ and the new scale $M$ which governs the deviations from TEGR so that in the limit $M\rightarrow\infty$ local Lorentz symmetry is restored and we recover TEGR. We can introduce St\"uckelberg fields $\lambda^a{}_b$ to restore the local Lorentz symmetry at all scales by introducing the replacement 
\be
	e^a{}_\mu\rightarrow \left(\exp\lambda/M\right)^a_{\ \, b} e^b{}_\mu. 
\ee	
In these variables, the leading order action for the tetrads is described by the TEGR action, while the St\"uckelberg fields will only enter in the higher order operators. If we assume a flat tetrad background, then its perturbations will be normalised with $\mpl$ and we can take a decoupling limit $M_P\to\infty$ where the tetrad perturbations completely decouple from the St\"uckelberg fields. In this decoupling limit, it is justified leaving only perturbations in $\lambda$ and focus in that sector.

\section{A perturbative treatment}

Let us consider the general Minkowski solution as perturbation around the trivial tetrad, or around $\Lambda=I$. Due to the preserved \emph{global} Lorentz invariance, however, any other constant tetrad might also be used. It is convenient to go for the Lorentz algebra instead of the Lorentz group. We write the usual exponential map as
\be
\label{exp}
	\Lambda = \exp(\lambda) = I + \lambda + \frac{1}{2} \lambda^2 + \frac{1}{3!} \lambda^3+\ldots \, ,
\ee
where $\lambda \in so(1,3)$. From now on, for the sake of perturbation theory, the indices would be raised and lowered by $\eta_{\mu\nu}$, and changed by the trivial tetrad $\delta^a_{\alpha}$, so that we will write $\lambda$ components as $\lambda_{\mu\nu} = -\lambda_{\nu\mu}$.  

The action takes the form
\be
\label{Lact}
	S=\int \dd^4 x \cdot f\left(2\partial^{\mu}(\Lambda^{-1}\partial_{\nu}\Lambda)^{\nu}{}_{\mu}\right),
\ee
and can be expanded in powers of $\lambda$ using
\be
\label{series}
	\Lambda^{-1}\partial \Lambda 
	= \int^{1}_{0} \dd u \, e^{-u \, \ad_\lambda} \,  \partial\lambda
	= \partial\lambda + \frac{1}{2} [\partial\lambda, \lambda] + \frac{1}{3!} [[\partial\lambda, \lambda], \lambda] + \frac{1}{4!}[[[\partial\lambda, \lambda], \lambda], \lambda] + \ldots\,,
\ee
where $\ad_\lambda X \equiv [\lambda, X]$. Using the commutativity of partial derivatives and the antisymmetry of the elements of the Lie algebra, the torsion scalar can be brought into a manifestly first derivative form
\bea
\label{TT}
	{\TT} 
&	= 2 \int^{1}_{0} \dd u \int^{u}_{0} \dd v \, [ e^{-u \, \ad_\lambda} \partial_{\nu} \lambda, e^{-v \, \ad_\lambda} \partial^{\mu} \lambda ]^{\nu}{}_{\mu} \\
&	=  [ \partial_{\nu} \lambda, \partial^{\mu} \lambda ]^{\nu}{}_{\mu} + \frac{2}{3}[ [\partial_{\nu} \lambda, \lambda], \partial^{\mu} \lambda ]^{\nu}{}_{\mu} + \frac{1}{3}[ \partial_{\nu} \lambda, [\partial^{\mu} \lambda,\lambda] ]^{\nu}{}_{\mu} + \ldots \, ,
\eea
thus, self-consistently, only first order derivatives of $\lambda$ will appear in the action. \B{Moreover, note that \eqref{TT} is linear in time derivatives of $\lambda$. So, the kinetic matrix corresponding to \eqref{Lact} has rank one as long as $f''(0) \neq 0$. Therefore, all six canonical momenta depend on one and the same combination of velocities implying that there are five primary constraints for the six canonical momenta. This suggests the existence of one extra mode.} 

We remark that the global Lorentz symmetry can be used to set $\lambda = 0$ at any fixed point and so only the leading term contributes at this point.\footnote{The global Lorentz transformation $\Lambda \to e^{\epsilon} \Lambda$, with $\epsilon$ a constant and antisymmetric, corresponds to a non-linear transformation in the Lie algebra due to the non-Abelian nature of the Lorentz group. The exact correspondence can be obtained using, e.g., the Baker–Campbell–Hausdorff formula. For infinitesimal transformations $\lambda \to \lambda + F(\ad_{\lambda}) \epsilon + \mathcal{O}(\epsilon^2)$, with $F(u) = u/(e^u-1)$. The non-linearity in $\lambda$ must be accounted for when working at higher orders in perturbation theory.} 
It follows that the leading term is sufficient to capture the smooth evolution of $\lambda$ in a sufficiently small neighbourhood of any given point regardless of the value of $\lambda$ at that point.

In this picture, $f(T)$ gravity can be interpreted as a metric theory with an antisymmetric field $\lambda_{\mu\nu}$ living on top of that. There is little difference from the covariant viewpoint obtained by expanding the inertial spin connection around the vanishing Weitzenb{\" o}ck one.

Since the linear term in the expansion of $f$ gives a surface term, the leading contribution is produced at the fourth order in $\lambda$,
\bea
\label{lowact}
	S
&	= \frac{ f''(0) }{2} \int \dd^4 x \,  \left( \lambda_{\rho}{}^{\nu,\mu} \lambda^{\rho} {} {}_{\mu,\nu} -\lambda_{\rho}{}^{\nu}{}_{,\nu} \lambda^{\rho \mu}{}_{,\mu} \right)^{2} \\
\eea
In order to separate space and time we will define variables representing rapidities and angles, $\lambda_{0i} \equiv \zeta_i$ and $\lambda_{ij} \equiv \epsilon_{ijk}\theta_k$. We obtain
\bea
\label{lowact_2}
	S
&	=  \frac{ f''(0) }{2} \int \dd^4 x
	\left[2\epsilon_{ijk}  \Big(\dot \zeta_i \theta_{k,j} + \dot \theta_i \zeta_{k,j}\Big) + \zeta_{i,i}^2 - \theta_{i,i}^2 - \zeta_{i,j}\zeta_{j,i} + \theta_{i,j}\theta_{j,i}\right]^2.
\eea
\B{We can explicitly observe a linear combination of velocities in the bracket, as was noted above, implying the existence of one extra mode.} The main lesson, that strongly coupled extra mode(s) appear at 4$^{th}$ order can be taken without entering into details of a Hamiltonian analysis.

The quadratic expression in the brackets can be diagonalized by introducing the complex field $X_{i} = \theta_i + i \zeta_i$,
\bea
\label{lowact_3}
	S
&	=  \frac{ f''(0) }{2} \int \dd^4 x
	\left[{\rm Re}\left(-2 i\epsilon_{ijk} \dot X_i X_{k,j} +X_{i,j}X_{j,i} - X_{i,i}^2 \right)\right]^2
\eea
Naively this suggests the existence of 3 extra modes, in accordance with~\cite{Li:2011rn}. Due to the divergence-like
structure of the kinetic term however, we expect only one dynamical dof to survive the constraints, thus our result does not contradict~\cite{Ferraro:2018tpu}. 

We remark that the first term in \eqref{lowact_3} implies a possible relation with certain electromagnetic theories as it has the same shape as the pseudoscalar $\epsilon^{\mu\nu\rho\sigma}F_{\mu\nu}F_{\rho\sigma} = 4\epsilon_{ijk} \dot A_i A_{k,j}$  in the Weyl gauge, $A_0 = 0$.

\section{Discussion}

In summary, we see that the extra modes appear only at the 4-th order in the action or, equivalently, at the 3-rd order in equations of motion. In order to study these modes, we have proposed a simplified framework based on the decoupling of the new dof on a flat metric. Although, from the divergence-like
structure of the kinetic term, we expect only one new mode, it would be premature to claim that this simple calculation shows the presence of{ \it only} one extra mode. 

On the more concrete side, we have actually seen the new mode alive, and now we know the associated energy scale: it comes with the dimension 8 operator $(\partial\lambda)^4$ scaling in the action. Seeing explicitly the additional propagating dof around the Minkowski background and establishing that it arises at the fourth order in perturbation theory may have dramatic consequences for the phenomenology of these theories since it clearly shows the aforementioned strong coupling. This extra mode does not appear on cosmological backgrounds either, posing a serious shortcoming of the cosmologies based on these theories. The problem can be understood in the usual manner, namely, the surface corresponding to cosmological solutions in phase space is singular and, hence, background solutions lying on that surface are prone to pathologies. In other words, if we start from a configuration close to this singular surface, the kinetic term of, at least one, propagating mode is arbitrarily small and thus it is prone to large effects from small disturbances. In practice, this means that the singular surface cannot be an attractor in the phase space and it will have unstable directions so that cosmological solutions cannot be perturbatively trusted. The problem in the flat background can be traced to the appearance of dynamical modes for the Lorentz St\"uckelberg fields that describe a 2-form field. This jeopardy has also been identified in other torsion-based theories like New GR \cite{Hayashi:1979qx} where the 2-form must be endowed a gauge symmetry at linear order around Minkowski \cite{Ortin:2015hya,Koivisto:2018loq} that is necessarily broken at higher orders \cite{Jimenez:2019tkx}, with the corresponding strong coupling. To some extent, the found pathology is similar to the one present in non-linear extensions of Gauss-Bonnet gravities. At the full non-linear order, it is possible to see that there is one extra scalar field (see e.g. the explicit construction in \cite{Kobayashi:2011nu}) that is absent from the linear spectrum around Minkowski. It is also interesting to notice the potential relation with the remnant Lorentz symmetries that are found around certain backgrounds. These accidental symmetries are arguably related to the evanescent  dof's from the linear spectrum so they can be used to diagnose strongly coupled background solutions.

Finally, $f(T)$ gravity can be considered a special case of scalar-torsion gravity~\cite{Cai:2015emx,Hohmann:2019gmt,Flathmann:2019khc,Emtsova:2019qsl} as it is equivalently described by the Lagrangian $\phi T - F(\phi)$, where $\phi$ an auxiliary field and $F$ is the Legendre transformation of $f$. In both cases the local Lorentz violation can be seen to occur due to an $\phi T$ coupling, which indicates that the extra dof's found here, and the potential stability issues associated with them, might be present in a much broader class of teleparallel theories. For example, our result gives a good reason to expect that the  ``teleparallel Horndeski" theories \cite{Bahamonde:2019shr,Bahamonde:2020cfv} contain more than one scalar degree of freedom in their spectrum.

\acknowledgements  JBJ acknowledges support from the  {\it Atracci\'on del Talento Cient\'ifico en Salamanca} programme and the MINECO's projects FIS2014-52837-P and FIS2016-78859-P (AEI/FEDER). TSK was funded by the Estonian Research Council grants PRG356 and MOBTT86, and by the European Regional Development Fund CoE program TK133 ``The Dark Side of the Universe". HV was supported by the European Regional Development Fund through the CoE program grant TK133, the Mobilitas Pluss grants MOBTP135, MOBTT5 and by the Estonian Research Council grant PRG803.

\bibliography{fTMink}

\end{document}